\documentclass[%
reprint,
footinbib,
 amsmath,amssymb,
]{revtex4-1}
\usepackage{comment}

\usepackage{graphicx}% Include figure files
\usepackage{dcolumn}% Align table columns on decimal point
\usepackage{bm}% bold math
\usepackage{amsmath, amssymb}

\usepackage{color}
\usepackage{xfrac}
\usepackage{ulem}
\begin{document}

%\preprint{APS/123-QED}

\title{Pattern of confined chemical garden controlled by injection speed}% Force line breaks with \\
%\thanks{A footnote to the article title}%

\author{Shu Wagatsuma}
\affiliation{%
Department of Applied Physics, Tokyo University of Science, Tokyo 125-8585, Japan
}%

\author{Takuro Higashi}
\affiliation{%
	Department of Applied Physics, Tokyo University of Science, Tokyo 125-8585, Japan
}%

\author{Ayumi Achiwa}
\affiliation{Department of Education, Aichi University of Education, Aichi 448-8542, Japan
}%

\author{Yutaka Sumino}
\email{ysumino@rs.tus.ac.jp}
\affiliation{%
Department of Applied Physics, Tokyo University of Science, Tokyo 125-8585, Japan
}%

%\collaboration{CLEO Collaboration}%\noaffiliation

\date{\today}% It is always \today, today,
             %  but any date may be explicitly specified

\begin{abstract}
Pattern of confined chemical garden was controlled by the speed of injected fluid, and their mechanism is discussed. A confined chemical garden system was constructed where an aqueous solution of cobalt chloride was injected into a cell filled with sodium silicate solution. The reaction of these two solutions resulted in the formation of precipitation. The viscosities of the prepared aqueous solutions were set to be similar in order to rule out the possibility of Saffman-Taylor instability. The injection front showed three distinctive patterns: algaes, shells, and filaments, which were dependent on injection speed. The injection pressure and the spatio-temporal pattern of the injected fluid were measured, and a significant increase in the injection pressure was observed when the filament pattern appeared, which indicated the existence of thin lubrication layer between the precipitation and the substrate. The filament pattern was further analyzed quantitatively, and the number of active filaments was determined to be proportional to the injection speed. A mathematical model was constructed that considered both the viscous effect from the thin luburiation layer and the Laplace pressure. This model successfully reproduced the characteristic filament dynamics.

%\begin{description}
%\item[Usage]
%Secondary publications and information retrieval purposes.
%\item[PACS numbers]
%May be entered using the \verb+\pacs{#1}+ command.
%\item[Structure]
%You may use the \texttt{description} environment to structure your abstract;
%use the optional argument of the \verb+\item+ command to give the category of each item. 
%\end{description}
\end{abstract}

% PACS, the Physics and Astronomy
                             % Classification Scheme.
%\keywords{Suggested keywords}%Use showkeys class option if keyword
                              %display desired
\maketitle

%\tableofcontents

\section{Introduction}
Placing a grain of a metal salt at the bottom of a sodium silicate solution results in the formation of a branching aggregate that grows spontaneously upward. This phenomenon is referred to as a chemical garden~\cite{01, oliver1, oliver02, oliver03}. The essential factor for the growth of a chemical garden is the formation of a gel made of precipitation of metal cation and anions such as silicate. When a grain of metal salt is placed in a sodium silicate solution, a membrane made of a gel, through which water can permeate, covers the salt. The concentration of salt in the solution inside and outside of the membrane is different, which results in osmotic pressure. The membrane expands due to this osmotic pressure and finally ruptures. In this way, a concentrated aqueous solution of metal salt is ejected from the membrane, which in turn is surrounded by the gel membrane. This process is thus repeated and results in a branch-like aggregate. 

Such chemical gardens are a manifestation of a coupling of flow and precipitation. Hydrodynamic flow is generated by osmotic pressure, and precipitation occurs when the metal salt and sodium silicate solution mix to form a gel. This same coupling of flow and precipitation especially under frictional environment is also relevant for geophysical conditions~\cite{15}, cellular motility~\cite{17}, as well as industrial application~\cite{Karol2003}. Such systems often contain various chemicals and/or biological agents, and thus a simplified physical view would be valuable for predicting the behavior of these systems. Therefore, it is of interest to produce a system where various spatio-temporal patterns appear according to the physical parameters that arise from a coupling of flow and precipitation under frictional environment~\cite{sumino01, sumino02}.

A simplified 2-dimensional (2D) chemical garden system was recently invented by Haudin et al.~\cite{PNAS, JPCC, 05}. The geometry of the system is constrained to be 2D by the use of a thin cell, known as a Hele-Shaw cell. One of two aqueous solutions of cobalt chloride and sodium silicate is placed in the cell, and the other solution is injected mechanically from the center of the cell. In this way, the hydrodynamic flow is imparted externally. Mixing of these two solutions results in the formation of precipitation~\cite{08}. Further interestingly, frictional environment is also introduced due to a Hele-Shaw cell. In their report, various spatio-temporal dynamics were identified, such as the formation of algaes, shells, and filaments, by changing the concentration and the chemical species. Therefore, the 2D chemical garden is a desirable candidate for exploration of the coupling of flow and precipitation under frictional environment. However, previous work has mainly used variation of the chemical parameters, while changes to physical parameters such as the flow rate and cell thickness have not been systematically examined.
Therefore, we have focused on the influence of physical parameters on 2D chemical garden formation. Cells with different thickness were filled with sodium silicate solution, into which cobalt chloride solutions were injected at various rates. These two solutions were set to have similar viscosities, so as to rule out the possibility of a typical Saffman-Taylor instability, i.e., viscous fingering~\cite{12,ParkHomsy1984}. 

\section{Experimental system}
Water was purified using a Millipore Milli-Q system. Cobalt chloride was purchased from Wako Pure Industries Ltd. Sodium silicate, provided by Fuji Chemical Co. Ltd., contained 9.4 wt\% sodium oxide and 29 wt\% silicon dioxide, which is almost equivalent to 6.5 mol/L of silicate in the aqueous phase. This sodium silicate solution was diluted to have 4.9 wt\% sodium oxide and 15 wt\% silicon dioxide, of which the viscosity was 3.8 mPa$\cdot$s, and this solution is referred to as water glass. An aqueous solution of 2 mol/L cobalt chloride was prepared, of which the viscosity was 2.4 mPa$\cdot$s, and this is referred to as the inner fluid.
\begin{figure}[tb]
	\centering
	\includegraphics[width=86mm, bb=0 0 235 240]{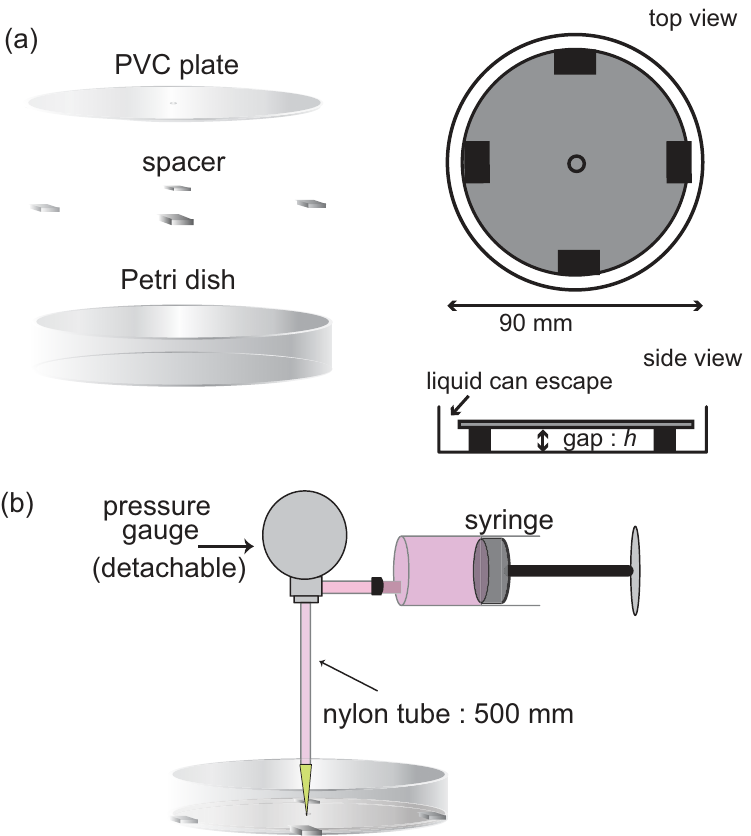}
	\caption{(a) Setup of Hele-Shaw cell made of poly(vinyl chrolide), PVC, plate with Petri dish. (b) Schematic illustration of experimental setup as a whole.}
	\label{schematic}
\end{figure}

\begin{figure}[tb]
	\centering
	\includegraphics[width=86mm, bb=0 0 262 164]{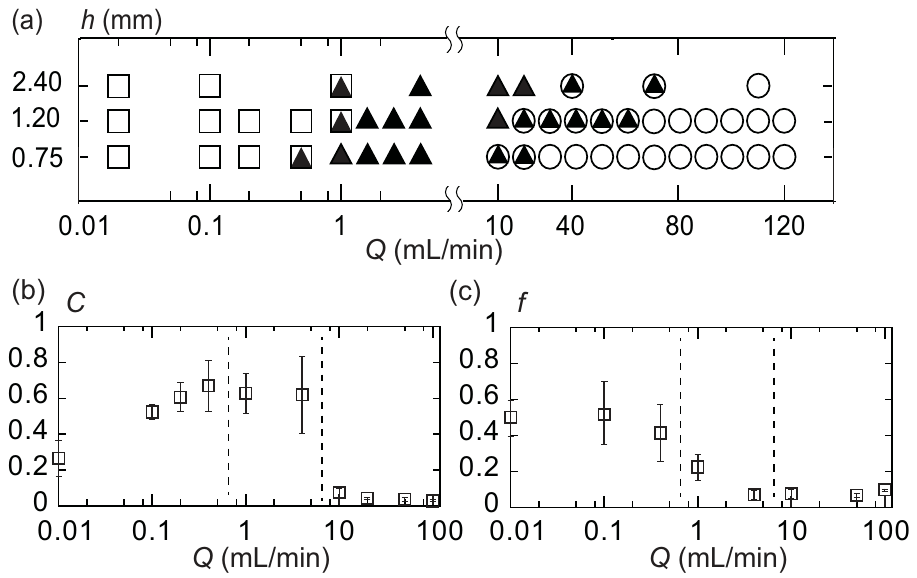}
	\caption{Phase diagram and order parameters $C$ and $f$, determined by the flow rate $Q$ and cell thickness $h$. (a) Phase diagram for the present system. Blank squares, filled triangles and blank circles represent algae (Fig.~\ref{flower}), shell (Fig.~\ref{shell}) and filament (Fig.~\ref{filament}) patterns. The overlapped symbols represent the coexistence of these patterns. The boundaries between patterns shift to higher $Q$ with increasing $h$. Horizontal axis is logarithmic up to $Q$=5, and liner for larger $Q$. (b,c) Dependence of the order parameters, (b) pattern circularity $C$, and (c) variance of areal velocity $f$, on $Q$, where $h$ is 0.75 mm. The filament pattern showed the lowest $C$, which indicates a branching shape, whereas the algae pattern showed the largest value of $f$, which indicates non-steady growth of the interface. The dotted lines represent the boundary where only shell patterns were observed. The errorbars correspond to $\pm$ standard deviation.}
	\label{phase}
\end{figure}

\begin{figure}[tb]
	\centering
	\includegraphics[width=86mm, bb=0 0 240 246]{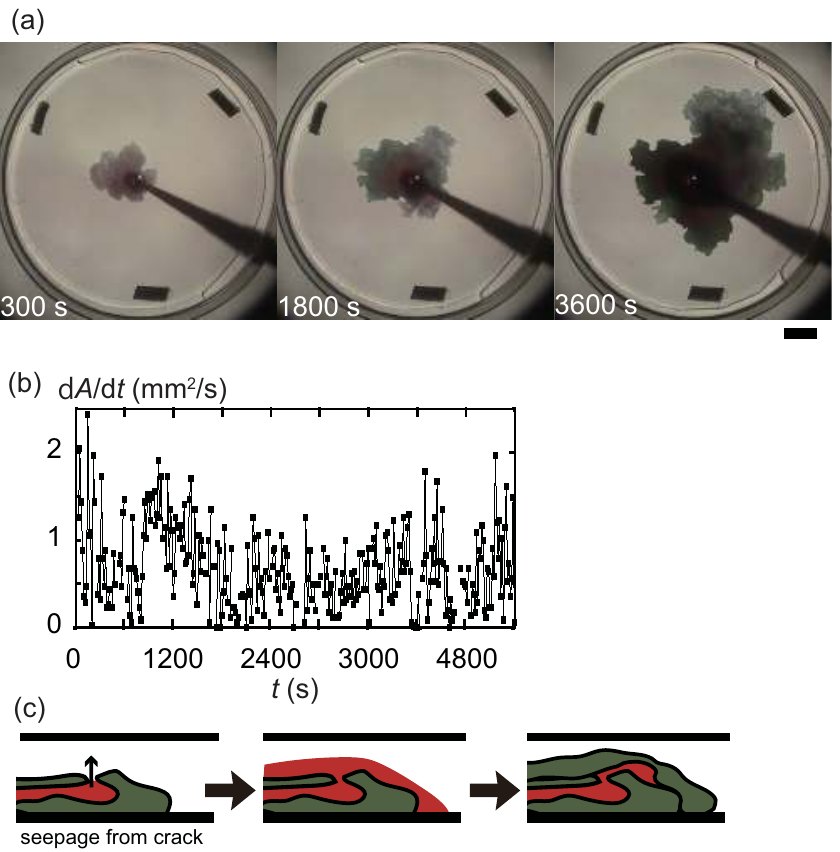}
	\caption{(a) Snapshots of algae pattern growth, where $Q$ = 0.01 mL/min. The interface was rough, and the color of the aggregate changed with time, which reflects the change in the hydration structure. Scale bar: 10 mm. (b) Time course of areal velocity, $dA/dt$, where a large fluctuation was observed. Such intensive fluctuation in $dA/dt$ indicates that the generated aggregate had a 3D structure. (c) Schematic illustration of the algae type aggregates in side view. Inner fluid leaks out from small fracture of precipitates. See also supplementary information for movies\cite{suppl}.}
	\label{flower}
\end{figure}

The experimental system consisted of a horizontal Hele-Shaw cell with gap width of $h =$ 0.75, 1.2 and 2.4 mm. The cells were constructed with 90 mm diameter polystyrene Petri dishes (1-8549-04, AS ONE Co.) and poly (vinyl chloride), PVC, cover plates (Sekisui Co., Ltd.) (Fig.\ref{schematic}(a)). There was a space around the cover plate to allow solution to escape. The Hele-Shaw cell was filled with water glass and the inner fluid was then injected from the center of the upper plate at various injection rates ($Q$ = 0.01 to 120 mL/min.). The central hole, whose diameter was 1 mm, and a 50 mL syringe (Henke Sass Wolf Co. Ltd) were connected with a nylon tube (Nihon Pisco Co., Ltd.; 2.5 mm internal diameter) (Fig.\ref{schematic}(b)). The syringe was installed on a syringe pump (CXF1010; ISIS Co., Ltd.). A pressure gauge (60X10KPA, Daiichi Keiki MFG. Co., Ltd; S010.1MP, Migishita Keiki MFG. Co., Ltd.) inserted between the injection point of the cell and the syringe pump was used to measure the pressure. The length of the tube from the branching point for the pressure gauge to the injection point of the cell was 500 mm. The pattern of injected fluid was measured from the bottom of the cell with a digital video camera, and analyzed using Image J software (NIH) ~\cite{17}. 

\begin{figure}[tb]
	\centering
	\includegraphics[width=86mm, bb=0 0 237 193]{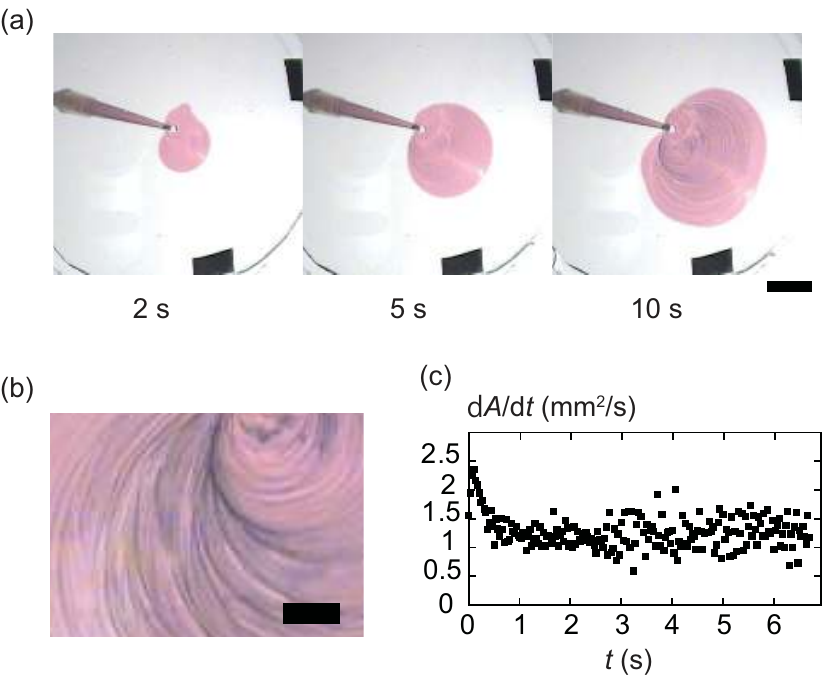}
	\caption{(a) Snapshots of shell pattern growth, where $Q$ = 4 mL/min. The advancing interface was smooth, and a periodic stripe pattern was observed in its trace. Scale bar: 10 mm. (b) Magnified image of the stripe pattern, for which the period was typically ca. 2 mm. Scale bar: 5 mm. (c) Time course of areal velocity, $dA/dt$. Except for an initial disturbance observed prior to 0.5 s, the areal velocity was almost constant. See also supplementary information for movies\cite{suppl}.}
	\label{shell}
\end{figure}

\begin{figure}[tb]
	\centering
	\includegraphics[width=86mm, bb=1 0 199 278]{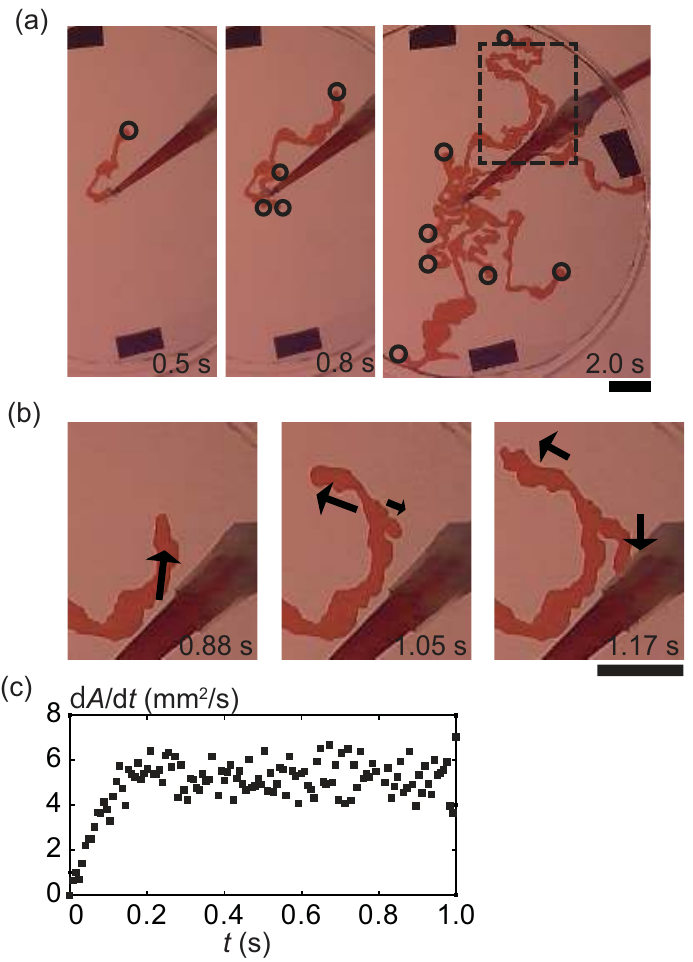}
	\caption{(a) Snapshots of filament pattern growth, where $Q$ = 30 mL/min. The circles correspond to the tips of the moving filaments. A box encircled by dashed line represents the close up region in (b). Scale bar: 10 mm. (b) Close up version of snapshots of filaments which showed splitting  of a tip, where one active filament became two. (c) Time course of areal velocity, $dA/dt$. Despite the complexity of the pattern, the areal velocity was almost constant. See also supplementary information for movies\cite{suppl}.}
	\label{filament}
\end{figure}

\section{Results and discussion}
Three distinctive patterns were obtained by varying $Q$ and $h$, as exemplified in Fig.~\ref{phase}(a). Algae, shell and filament patterns were observed (Figs.~\ref{flower}-\ref{filament}), as reported previously by Haudin et al.~\cite{PNAS}. An increase of $Q$ leads to a pattern transition from algae, to shell, and to filament. In contrast to the report in~\cite{PNAS}, where the chemical composition was changed, variation of the physical parameters $Q$ and $h$ resulted in three different patterns. Furthermore, these patterns were successfully characterized according to the order parameters; the circularity $C$, and the variance of areal velocity, $f$. $C$ is defined as $C=4 \pi A/\Pi^2$, where $A$ and $\Pi$ are the area and the perimeter of the injected fluid, respectively. Thus, in typical notation, $C$ is 1 for a circle and less than 1 for a rough pattern. $C$ was measured when $t=\tau$, i.e., when 1 mL of the inner fluid was injected. $f$ is defined as $f=\left(\langle \left(\frac{dA}{dt}\right)^2 \rangle-\langle\frac{dA}{dt}\rangle^2 \right)/ \langle \frac{dA}{dt} \rangle^2$, where $\langle \rangle$ denotes the temporal average over $t=0$ to $\tau$. Figures~\ref{phase}(b) and (c) show $C$ and $f$ when $h$ = 0.75 mm as a function of $Q$. $C$ was small and became close to zero when the filament pattern appeared (Fig.~\ref{phase}(b)). Thus, $C$ is a relevant parameter to distinguish a filament pattern from a shell pattern. A small decrease in $C$ is noted for an algae pattern compared with a shell pattern; however, the difference becomes much more significant when a change in $f$ is observed, as shown in Fig.~\ref{phase}(c). The algae pattern was characterized by random intermittent growth, which resulted in high $f$. The shell pattern also shows a relatively high $f$ compared to the filament pattern. The boundary between the algae and shell patterns seems to be continuously changed. The algae pattern was characterized by a much higher $f$, and thus $f$ is useful for differentiating algae from shell patterns. The coexistence of shell and filament patterns was also observed, not only in different trials with the same parameters, but also within the same trial. Such transitioning of patterns became more frequent when the gap size $h$ was increased; i.e., the boundary between each pattern became unclear. 

\begin{figure}[tb]
	\centering
	\includegraphics[width=86mm, bb=1 0 237 190]{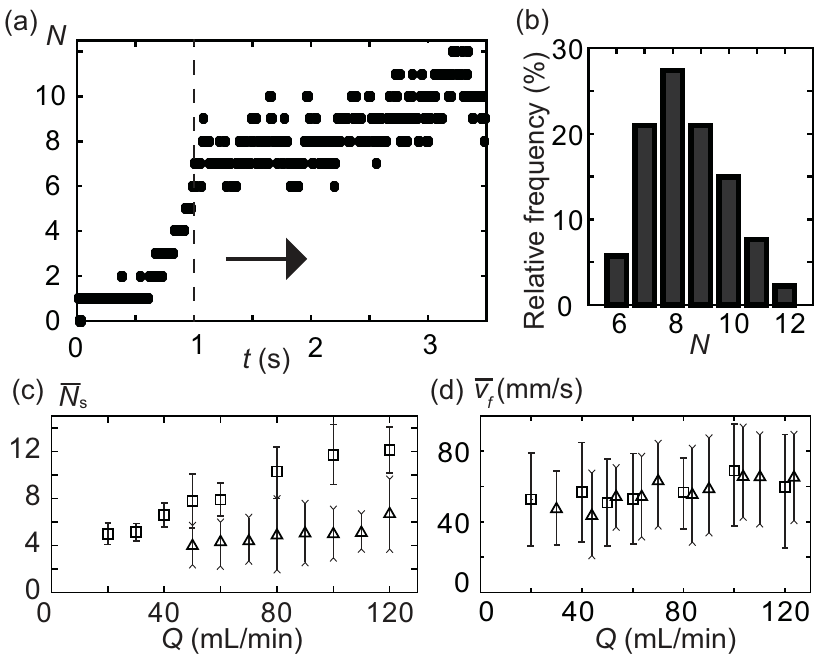}
	\caption{(a) Dependence of the number of active filaments $N$ on $t$. $N$ become saturated over time. The saturated value ${N}_{s}$ was measured after $t=1$ s, as indicated by the dashed line. (b) Histogram of relative frequency of $N$ observed after $t=1$ s, as indicated by the dashed line in (a). (c,d) Dependence of (c) $\bar{N}_{s} $ and (d) the filament velocity $\bar{v}_{f}$ on $ Q $, where data for $h$=0.75 mm and 1.2 mm are denoted by squares and triangles, respectively. $ \bar{N}_{s} $ increased steadily with $Q$, whereas the average $\bar{v}_{f}$ was almost constant, irrespective of $Q$ and $h$. The errorbars correspond to $\pm$ standard deviation.}
	\label{filament2}
\end{figure}

For small $Q$, 3D aggregates were formed at the front of the injected fluid, and the inner fluid was ejected from a small rupture in the aggregate, as shown in Figure~\ref{flower}(a). The top-view of the front shape had fractal-like millimeter-scale roughness, and expansion of the front was irregular and characterized by fluctuation of the 2D areal velocity $\frac{dA}{dt}$, as shown in Fig.~\ref{flower}(b). Significant fluctuation in the areal velocity $\frac{dA}{dt}$, and hence large $f$, despite a constant injection rate $Q$, indicated that the generated aggregate did not span the entire gap, i.e., the aggregate had a 3D structure (Fig.~\ref{flower}(c)). Furthermore, due to slow front motion (less than 10 $\mu$m/s), the aggregate changed color during growth over time, which suggests a change in the chemical composition within the precipitation~\cite{PNAS,colour}.

In the intermediate $Q$ range, shell patterns were observed. The shell pattern is characterized by a repeated extension of a smooth injection front (Fig.~\ref{shell}(a)). This resulted in a trace with a shell-like texture, for which the period was typically 2 mm, as shown in Fig.~\ref{shell}(b). Even though the extension of the rim had temporal oscillations, $\frac{dA}{dt}$ was almost constant overall (Fig. \ref{shell}(c)). The constant increase in the injected area, in addition to the small $f$ (Fig.~\ref{phase}(c)), suggest that the generated aggregate spanned the entire gap. 

A similar shell structure was reported in the case when water glass was injected into a cobalt chloride solution~\cite{JPCC}, but not in the opposite case. The difference between the present study and that reported in~\cite{JPCC} can be attributed to the concentration difference (higher cobalt chloride concentration and different composition of water glass), in addition to a different injection speed. As reported in~\cite{JPCC}, the coexistence of shell and filament patterns at the marginal value of $Q$ was also observed in the present study.

When $Q$ was sufficiently large, the front of the injected fluid branched to form multiple fronts, i.e., a filament pattern (Fig.~\ref{filament}(a)). The front exhibited irregular meandering and splitting (Fig.~\ref{filament}(b)), in contrast to typical viscous fingering due to the Saffman-Taylor instability~\cite{12, ParkHomsy1984}. However, the fluctuation in $\frac{dA}{dt}$ was small, as shown in Fig.~\ref{filament}(c), which resulted in small $f$. 

To quantify the branching dynamics, the number of active (mobile) filaments $N$ was counted during filament growth over time. To enumerate the number of active filaments, difference images were constructed from snapshots separated by a fixed time, $\Delta t$. Here, $\Delta t(Q,h)$ is the time span to have the same total areal increment $\Delta A$ as a function of $Q$ and $h$. In this study, $\Delta A \sim 27$ mm$^2$ was selected. For example, if $Q$ = 30 mL/min and $h$=0.75 mm, then $\Delta t$ = 0.04 s. In this way, the advancing front is extracted and the fronts of areas larger than 1.3 mm$^{2}$ are counted as actively moving filaments. In addition to $N$, the speed of the moving front $v_f$ was also measured, which represents the speed of the filaments.

Figure~\ref{filament2}(a) shows a typical time course for the number of active filaments $N$; the number of moving fronts $N$ initially increased with time and then satuated at finite value, $N_s$. This can be also observed by the peak in the histogram of the observed number of active filaments (Fig.~\ref{filament2}(b)). For $h$ = 0.75 and 1.2 mm, $\bar{N}_s$ and $\bar{v}_f$ were obtained from four independent measurements for each parameter with $Q$, where a bar denotes the temporal average. $\bar{N}_s$ was proportional to $Q$ (Fig.~\ref{filament2}(c)), whereas the front velocity $\bar{v}_f$ was almost the same for all $h$ and $Q$, with large fluctuations shown in Figure~\ref{filament2}(d).

\begin{figure}[tb]
	\centering
	\includegraphics[width=86mm, bb=1 0 243 102]{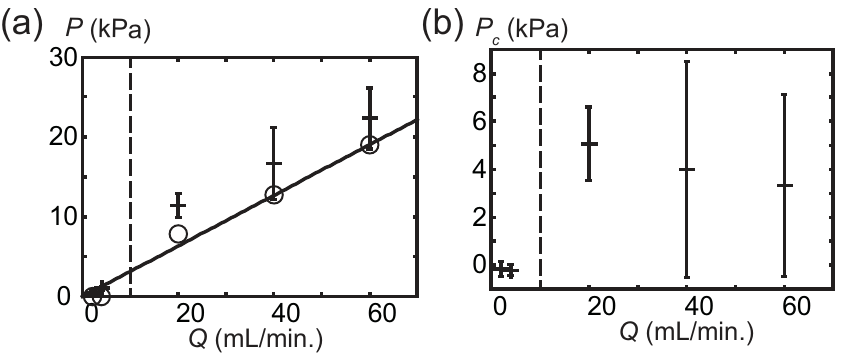}
	\caption{(a) Measured pressure $P$ with respect to injection rate $Q$. Blank circle represents $P_w$ when inner fluid was injected into pure water, where there were no precipitatoin reaction occurred. Solid line represents the fit for these data, whose slope is obtained as $\Xi =$0.32 kPa/(mL/min.). Cross with error bar represents $P$ when inner fluid was injected into water glass where precipitation reaction occurs. Three independent measurement was conducted, and oressyre was measured when 1 mL of inner fluid was injected. (b) Assumed pressure at the center of the cell $P_c$ when precipitation occurred. $P_c$ was obtained by substracting the pressure required for injecting water (assumed from fitted value), $\Xi Q$, from measured pressure, $P$. The data, $Q <10$ mL/min. (denoted by a vertical dashed line) corresponds to shell pattern, whereas the other data corresponds to filament pattern. The errorbars correspond to $\pm$ standard deviation.}
	\label{press}
\end{figure}

The pressure measurements were also conducted while observing the pattern development. In order to obtain friction coefficient of cell without precipitation reaction $\Xi$, we filled the same cell with a pure water, and injected inner fluid with various rate $Q$. The result of this measurement is shown in Figure~\ref{press}(a) with blank circle. Fitting these observed data with $P=\Xi Q$, we obtained $\Xi = 0.32 \pm 0.01$ kPa/(mL/min.). This $\Xi$ is high if we estimate it with nominal geometrical setup\cite{pressure, pressure2}. We infer that this discrepancy is due to the focusing flow just before the injection point and/or relatively narrow gap at the injection point. We, then, measured the pressure while injecting inner fluid, cobalt chloride solution, into the water glass. Here we measured the pressure value when 1 mL of inner fluid was injected. The pressure observed with $Q = 0.01$ mL/min, with alga pattern, was 150 Pa. The increased pressure due to the precipitation reaction, $P_c$, was is obtained by subtracting $\Xi Q =$ 3.2 Pa, and is about 150 Pa. This is high for the flow rate $Q$\cite{pressure2}, indicating flow within a narrow channel of precipitation is important for the algae pattern (Fig.~\ref{flower}(c)). In case of the shell and filament, we conducted systematic measurements shown in Figure~\ref{press}(a). For each $Q$, three independent measurements were conducted. Shell (filament) patterns were observed when $Q$ was smaller (larger) than 10 mL/min indicated as dashed line in Fig.~\ref{press}(a, b) . Coexistence of patterns were observed around 10 mL/min., so these data were not plotted in Figure~\ref{press}(a). Interestingly, deviation from the value without precipitation reaction were noted in the case with filament pattern, and is clearly seen with $P_c = P- \Xi Q$. From this, we can see that the pressure required for generates the shell pattern is of the order of 100 Pa, whereas the pressure required to have the filament pattern is of the order of kPa.

\section{Mathematical Modeling}
From the pressure measurement, the key factor for an algae pattern with a 3D structure is the formation of narrow fractures. The pressure required to advance the liquid increases inside of such narrow fractures. This dynamics involve microscopic cracking of the 3D aggregate, coupled with sequential chemical reactions indicated by the color change of the precipitate. To understand the mechanism for the formation of an algae pattern requires a detailed study of the sequential chemical reactions and shear effects, and will thus be the subject of future study.

In contrast, the injection pressure required for a shell pattern was almost the same as that required for algae pattern despite of high $Q$. 
A further increase in $Q$ resulted in an increase in the injection pressure to about kPa. At first consideration, this appears to be contradictory because an increase in $Q$ leads to a shorter reaction time and the aggregate should be less solidified. Therefore, the results suggest that the main factor influencing the dissipation of the filament pattern is different from those of algae and shell pattern. We speculate that this paradox is due to thin lubrication layer between the precipitation and the substrate that appears at the front of the injected fluid. Furthermore, we noted that the fluid inside of the filament did not precipitate at all except for their advancing boundary from the inspection of cells after the experiments. Therefore, we conclude that the generated precipitation should follow the advancing front.

\begin{figure}[tb]
	\centering
	\includegraphics[width=86mm, bb=0 2 220 69]{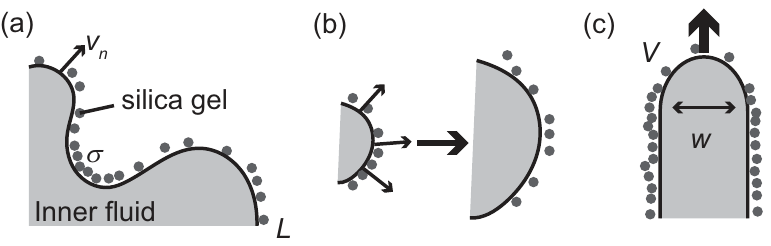}
	\caption{(a) Schematic representation of the boundary dynamics model. The model is based on movement of the 2D boundary $L$. Typical outward normal velocity is represented by $v_n$ and the local density of precipitation is represented by $\sigma$. (b) Decrease in $\sigma$ due to the curvature effect. (c) Simplified situation where the advancing tip moves with velocity $V$ and width $w$.}
	\label{model}
\end{figure}

Based on our speculation, a mathematical model was developed for the most interesting filament pattern. This modeling involved extension of the 2D boundary dynamics model developed for the drying processes of wet granular materials~\cite{02}. Here, the developed model includes the effect of precipitation as an obstacle to movement of the interface. The water glass/inner fluid interface is assumed to be a moving boundary $L$. 
The variable $\sigma$ was adopted to indicate the local density of precipitation at the moving boundary, for which the normal velocity is denoted by $v_n$ (Fig.~\ref{model}(a)). The precipitation does not diffuse; therefore, the dynamics of $\sigma$ can be determined by the geometrical effect, which is given by:
\begin{equation}\label{sim:eq1}
	\frac{\partial \sigma}{\partial t}= a-\kappa \sigma v_n.
\end{equation}
$\sigma$ increases with time due to reaction between the water glass and inner fluid. Here, the first term corresponds to the reaction, and $\sigma$ is considered to increase linearly with time, and is thus assigned the coefficient $a$ for simplicity. The second term corresponds to the geometrical factor, and $\kappa$ is the curvature of $L$. The unit length in the moving interface increases to be $1+\kappa v_n$ with each unit of time; thus, the density of precipitation at the interface effectively decreases (Fig.~\ref{model}(b))~\cite{02}.

The motion of the interface is given by
\begin{equation}\label{sim:eq2}
	v_n= f(\sigma)(\Delta p -\gamma \kappa),
\end{equation}
where $f(\sigma)$ represents the mobility of the interface under the influence of thin lubrication layer between precipitation and the substrate, and $f(\sigma)$ is modeled by: 
\begin{equation}\label{eq_f}
	f(\sigma)=\left\{ 
	\begin{matrix}
		\xi (\sigma_s-\sigma) & 0 \leq\sigma \leq \sigma_s, \\
		0 & \sigma > \sigma_s. 
	\end{matrix}
	\right.
\end{equation}

The interface is stopped by an increase in $\sigma$, and this effect is taken as the function $f$, where $v_n$ becomes 0 when the precipitation density $\sigma$, reaches $\sigma_s$. The interface is pushed by the pressure difference across the boundary $L$, denoted as $\Delta p$, whereas it is hindered by the Laplace pressure $\gamma \kappa$. Thus, the complete dynamics of the interface can be described by Eqs. (\ref{sim:eq1}) and (\ref{sim:eq2}). Interestingly, the proposed boundary model can be a differential form of the model previously proposed for a spiral pattern~\cite{PNAS}.

To elucidate the essential features of this model, a situation is considered where the curvature $\kappa=2/w$ and the normal velocity $v_n=V$. This corresponds to the tip of a steadily advancing filament, of which the velocity and width are $V$ and $w$, respectively (Fig.~\ref{model}(c)). In this case Eqs. (\ref{sim:eq1}) and (\ref{sim:eq2}) are expressed as: 
\begin{equation}\label{eq_sig}
\frac{\partial \sigma}{\partial t}=a -\frac{2\sigma V}{w},
\end{equation}
and
\begin{equation}\label{eq_V}
V=\xi ( \sigma_s - \sigma )\left(\Delta p- \frac{2\gamma}{w} \right).
\end{equation}

By inserting Eq. (\ref{eq_V}) into Eq.(\ref{eq_sig}), the steady-state condition, $\partial \sigma/ \partial t = 0$, leads to:
\begin{equation}\label{eq_sig0}
\left(\sigma -\frac{\sigma_s}{2} \right)^2-\frac{\sigma_s^2}{4}+\frac{aw}{2\xi\left(\Delta p-2\gamma/w\right)}=0.
\end{equation}
To have a real root for $\sigma$, the condition for $\Delta p$ is obtained as:
\begin{equation}\label{ineqDeltap}
	\Delta p \geq \frac{2\gamma}{w}+\frac{2aw}{\xi \sigma_s^2}.
\end{equation}
The first term in Eq.~(\ref{ineqDeltap}) corresponds to the effect of the Laplace pressure, so that the increase of the filament size $w$ leads to a smaller injection pressure being required to overcome the Laplace pressure. On the other hand, the second term results from the geometrical effect that reduces the local concentration of the precipitation, which will be smaller when $w$ is smaller. Thus, there are optimal conditions for filament growth under minimum pressure. This minimum pressure can be easily calculated from:
\begin{eqnarray}
\Delta p&\geq& \frac{2\gamma}{w}+\frac{2aw}{\xi \sigma_s^2} \nonumber \\
%&=&\frac{2}{\sigma_s} \sqrt{\frac{a\gamma}{\xi}} \left( \frac{\sigma_s}{w}\sqrt{\frac{\gamma \xi}{a}}+ \frac{w}{\sigma_s} \sqrt{\frac{a}{\gamma \xi}}\right) \nonumber \\
&\geq& \frac{4}{\sigma_s} \sqrt{\frac{a\gamma}{\xi}}, \label{ineqDeltap2}
\end{eqnarray}
where equivalency is attained if and only if $w=w_0=\sigma_s \sqrt{\gamma \xi/a}$.
%\begin{equation}
%w=w_0=\sigma_s \sqrt{\frac{\gamma \xi}{a}}.
%\end{equation}
When equivalency is fulfilled, the pressure difference $\Delta p$ takes the minimum value $\Delta p_0= \frac{4}{\sigma_s} \sqrt{\frac{a\gamma}{\xi}}$. Furthermore, we have $\sigma = \sigma_0= \sigma_s/2$, and $V=V_0=\sqrt{a \gamma \xi}$. In summary, Eqs. (\ref{eq_sig}) and (\ref{eq_V}) for the tip of an active filament have a minimum $\Delta p = \Delta p_0$ for the steady state. Under this condition, the advance of one filament will require a flow rate of $Q_m = V_0 w_0 h = \sigma_s \gamma \xi h$. It should be noted that $Q_m$ is dependent only on the geometrical parameter $h$ and the physical properties $\sigma_s, \gamma$, and $\xi$.

The existence of a minimum pressure difference for the growth of filaments explains the experimental observations for the filament pattern. 
Initially, when the inner fluid is injected into a cell, the number of filaments increases due to fluctuation of the interface as a result of inherent noise in the experimental system. At this stage, the injection pressure can be higher than $\Delta p_0$. However, as the number of active filaments increases, the pressure becomes small, and finally reaches $\Delta p_0$. If the number of active filaments increases further, then the pressure becomes smaller than $\Delta p_0$, i.e., advance of the filaments is halted. In this way, the number of active filaments becomes saturated, so that all the active filaments fulfill the minimum pressure condition, $\Delta p_0$. At this stage, each filament requires $Q_m$ for a tip to advance. Thus, the saturated number of active filaments $N_s$, can be obtained by $N_s=Q/Q_m$. Therefore, as long as a filament pattern appears, the number of active filaments is proportional to the injection rate $Q$, whereas the rate of advance of the active filament is determined by $V_0$. 
The analysis of filament patterns revealed that the number of active filaments was saturated at $\bar{N}_s$, where $\bar{N}_s$ was almost proportional to the injection rate of the inner fluid, $Q$. Furthermore, the average velocity $\bar{V}_f$ did not change with $Q$. 

The initial increase in the number of active filaments requires further discussion. Our preliminary numerical calculations for the boundary dynamics showed that the model tended to show tip-splitting, which is reported as double spiral formation in~\cite{PNAS}. We consider that this tip splitting is one of the main factors for the increase in the filament number. In addition to the pressure effect and inhibition of the boundary collision, the $N_s$ dependence on $Q$ can be reproduced. Analysis of the numerical calculations will be discussed elsewhere~\cite{future1}.
  
Filament patterns have been often observed in various systems where flow and precipitation are coupled~\cite{Belmonte1, Nagatsu1}, such as in biological systems~\cite{cell}. Saturation of the number of actively moving filaments was also noted in~\cite{Belmonte1}. In this sense, the proposed boundary model can be a simple mathematical model to determine the fingering patterns generated by the coupling of flow and precipitation within a confined geometry. Active stress generation can also be included in the dynamics, such that spontaneous droplet deformation due to aggregate formation could also be modeled~\cite{sumino01, sumino02}.

Comparing numerical value by Eqs. (\ref{ineqDeltap}) and (\ref{ineqDeltap2}), which connects surface tension $\gamma$ and the size of finger $w$ with the injection pressure $\Delta p$. Assuming the observed branch size corresponds to the case with minimum pressure, i.e. $w = w_0$, we have $\Delta p_0 = 4 \gamma /w_0$. Inserting the typical values, $\Delta p_0$ = 1 kPa and $w$ = 1 mm, we obtain $\gamma \sim$ few 100 mN/m. This estimated value of interfacial tension $\gamma$ is high for ordinary liquid whose typical values are of the order of 10 mN/m, but close to the surface energy of glass whose typical value is of the order of 100 mN/m\cite{Rhee1977}. Thus, the surface tension $\gamma$ described in the model might correspond to the one of generated precipitates.

Finally, we would like to comment on the possible implication of our result on chemical grouting. Chemical grouting is a common industrical technique to stabilize soil by injecting precipitating solution~\cite{Karol2003}. As mentioned, chemical grouting share essential physics with our system: flow, precipitation, and frictional envrionment, i.e. flow is governed by Darcy's law. Our result indicates that the pressure as well as the pattern of precipitates drastically altered depending on the injection rate of solutions. We believe further analysis of our result will render relevant inforation for chemical grouting, where in situ measurement of spatio-temporal pattern is, in principle, impossible for the presence of soil.

\section{Conclusions}
In this study, various patterns generated due to a coupling of flow and precipitation in a Hele-Shaw cell were investigated. Variation of physical control parameters, such as the injection rate $Q$ and the thickness of the cell $h$ was examined. Three distinctive patterns were noted: algaes, shells and filaments. These patterns were well classified by the circularity $C$ and by the variance of areal velocity $f$. The large $f$ determined for algae patterns suggests that these patterns possess 3D structure, even under a confined quasi-2D condition. The filament pattern has small $C$, which indicates filament growth. Detailed analysis of the dynamics of the filament pattern indicated that the number of actively growing filaments becomes saturated with time at $\bar{N}_s$, which was almost proportional to the injection rate $Q$. The velocity of the growing filament $\bar{v}_f$ was not affected by the change in $Q$. A larger cell thickness $h$ resulted in smaller $\bar{N}_s$ but almost the same $\bar{v}_f$. Measurement of the injection pressure revealed the smallest injection pressure at the injection point for a shell pattern, which can be explained by the simple viscosity effect of the solution. These results suggest that a different mechanism for dissipation is dominant for the filament pattern. 

Based on these observations, a 2D boundary dynamics model was constructed with existence of thin lubrication layer between the precipitation and the substrate at the growth front. The model predicts the optimal size of a filament with minimum pressure required for the steady state. Further speculation led to a successful explanation for the filament pattern. Both the experimental system and the proposed mathematical model can serve as a simple example for the coupling of flow and precipitation, by which various spatio-temporal patterns are generated that reflect those observed in geophysical, biological, as well as engineering systems.

This work was supported by JSPS KAKENHI Grant JP16K13866 and JSPS KAKENHI Grant JP16H06478. This work was also partially supported by a JSPS Bilateral Joint Research Program between Japan and the Polish Academy of Sciences, and the Cooperative Research of ``Network Joint Research Center for Materials and Devices'' with Hokkaido University (No. 20161033).

Finally, the authors would like to thank Fuji Chemical Co. Ltd. for kindly providing sodium silicate samples.

%\bibliography{refs}% Produces the bibliography via BibTeX.

\begin{thebibliography}{25}%
	\makeatletter
	\providecommand \@ifxundefined [1]{%
		\@ifx{#1\undefined}
	}%
	\providecommand \@ifnum [1]{%
		\ifnum #1\expandafter \@firstoftwo
		\else \expandafter \@secondoftwo
		\fi
	}%
	\providecommand \@ifx [1]{%
		\ifx #1\expandafter \@firstoftwo
		\else \expandafter \@secondoftwo
		\fi
	}%
	\providecommand \natexlab [1]{#1}%
	\providecommand \enquote  [1]{``#1''}%
	\providecommand \bibnamefont  [1]{#1}%
	\providecommand \bibfnamefont [1]{#1}%
	\providecommand \citenamefont [1]{#1}%
	\providecommand \href@noop [0]{\@secondoftwo}%
	\providecommand \href [0]{\begingroup \@sanitize@url \@href}%
	\providecommand \@href[1]{\@@startlink{#1}\@@href}%
	\providecommand \@@href[1]{\endgroup#1\@@endlink}%
	\providecommand \@sanitize@url [0]{\catcode `\\12\catcode `\$12\catcode
		`\&12\catcode `\#12\catcode `\^12\catcode `\_12\catcode `\%12\relax}%
	\providecommand \@@startlink[1]{}%
	\providecommand \@@endlink[0]{}%
	\providecommand \url  [0]{\begingroup\@sanitize@url \@url }%
	\providecommand \@url [1]{\endgroup\@href {#1}{\urlprefix }}%
	\providecommand \urlprefix  [0]{URL }%
	\providecommand \Eprint [0]{\href }%
	\providecommand \doibase [0]{http://dx.doi.org/}%
	\providecommand \selectlanguage [0]{\@gobble}%
	\providecommand \bibinfo  [0]{\@secondoftwo}%
	\providecommand \bibfield  [0]{\@secondoftwo}%
	\providecommand \translation [1]{[#1]}%
	\providecommand \BibitemOpen [0]{}%
	\providecommand \bibitemStop [0]{}%
	\providecommand \bibitemNoStop [0]{.\EOS\space}%
	\providecommand \EOS [0]{\spacefactor3000\relax}%
	\providecommand \BibitemShut  [1]{\csname bibitem#1\endcsname}%
	\let\auto@bib@innerbib\@empty
	%</preamble>
	\bibitem [{\citenamefont {Barge}\ \emph {et~al.}(2015)\citenamefont {Barge},
		\citenamefont {Cardoso}, \citenamefont {Cartwright}, \citenamefont {Cooper},
		\citenamefont {Cronin}, \citenamefont {De~Wit}, \citenamefont {Doloboff},
		\citenamefont {Escribano}, \citenamefont {Goldstein}, \citenamefont {Haudin},
		\citenamefont {Jones}, \citenamefont {Mackay}, \citenamefont {Maselko},
		\citenamefont {Paganoo}, \citenamefont {Pantaleone}, \citenamefont {Russell},
		\citenamefont {Sainz-Diaz}, \citenamefont {Steinbock}, \citenamefont {Stone},
		\citenamefont {Tanimoto},\ and\ \citenamefont {Thomas}}]{01}%
	\BibitemOpen
	\bibfield  {author} {\bibinfo {author} {\bibfnamefont {L.~M.}\ \bibnamefont
			{Barge}}, \bibinfo {author} {\bibfnamefont {S.~S.~S.}\ \bibnamefont
			{Cardoso}}, \bibinfo {author} {\bibfnamefont {J.~H.~E.}\ \bibnamefont
			{Cartwright}}, \bibinfo {author} {\bibfnamefont {G.~J.~T.}\ \bibnamefont
			{Cooper}}, \bibinfo {author} {\bibfnamefont {L.}~\bibnamefont {Cronin}},
		\bibinfo {author} {\bibfnamefont {A.}~\bibnamefont {De~Wit}}, \bibinfo
		{author} {\bibfnamefont {I.~J.}\ \bibnamefont {Doloboff}}, \bibinfo {author}
		{\bibfnamefont {B.}~\bibnamefont {Escribano}}, \bibinfo {author}
		{\bibfnamefont {R.~E.}\ \bibnamefont {Goldstein}}, \bibinfo {author}
		{\bibfnamefont {F.}~\bibnamefont {Haudin}}, \bibinfo {author} {\bibfnamefont
			{D.~E.~H.}\ \bibnamefont {Jones}}, \bibinfo {author} {\bibfnamefont {A.~L.}\
			\bibnamefont {Mackay}}, \bibinfo {author} {\bibfnamefont {J.}~\bibnamefont
			{Maselko}}, \bibinfo {author} {\bibfnamefont {J.~J.}\ \bibnamefont
			{Paganoo}}, \bibinfo {author} {\bibfnamefont {J.}~\bibnamefont {Pantaleone}},
		\bibinfo {author} {\bibfnamefont {M.~J.}\ \bibnamefont {Russell}}, \bibinfo
		{author} {\bibfnamefont {C.~I.}\ \bibnamefont {Sainz-Diaz}}, \bibinfo
		{author} {\bibfnamefont {O.}~\bibnamefont {Steinbock}}, \bibinfo {author}
		{\bibfnamefont {D.~A.}\ \bibnamefont {Stone}}, \bibinfo {author}
		{\bibfnamefont {Y.}~\bibnamefont {Tanimoto}}, \ and\ \bibinfo {author}
		{\bibfnamefont {N.~L.}\ \bibnamefont {Thomas}},\ }\href@noop {} {\bibfield
		{journal} {\bibinfo  {journal} {Chem. Rev.}\ }\textbf {\bibinfo {volume}
			{115}},\ \bibinfo {pages} {8652} (\bibinfo {year} {2015})}\BibitemShut
	{NoStop}%
	\bibitem [{\citenamefont {Thouvenel-Romans}\ and\ \citenamefont
		{Steinbock}(2003)}]{oliver1}%
	\BibitemOpen
	\bibfield  {author} {\bibinfo {author} {\bibfnamefont {S.}~\bibnamefont
			{Thouvenel-Romans}}\ and\ \bibinfo {author} {\bibfnamefont {O.}~\bibnamefont
			{Steinbock}},\ }\href {\doibase 10.1021/ja0298343} {\bibfield  {journal}
		{\bibinfo  {journal} {Journal of the American Chemical Society}\ }\textbf
		{\bibinfo {volume} {125}},\ \bibinfo {pages} {4338} (\bibinfo {year}
		{2003})},\ \bibinfo {note} {pMID: 12670257},\ \Eprint
	{http://arxiv.org/abs/http://dx.doi.org/10.1021/ja0298343}
	{http://dx.doi.org/10.1021/ja0298343} \BibitemShut {NoStop}%
	\bibitem [{\citenamefont {Makki}\ \emph {et~al.}(2012)\citenamefont {Makki},
		\citenamefont {Roszol}, \citenamefont {Pagano},\ and\ \citenamefont
		{Steinbock}}]{oliver02}%
	\BibitemOpen
	\bibfield  {author} {\bibinfo {author} {\bibfnamefont {R.}~\bibnamefont
			{Makki}}, \bibinfo {author} {\bibfnamefont {L.}~\bibnamefont {Roszol}},
		\bibinfo {author} {\bibfnamefont {J.~J.}\ \bibnamefont {Pagano}}, \ and\
		\bibinfo {author} {\bibfnamefont {O.}~\bibnamefont {Steinbock}},\ }\href
	{\doibase 10.1098/rsta.2011.0378} {\bibfield  {journal} {\bibinfo  {journal}
			{Philosophical Transactions of the Royal Society A: Mathematical,Physical and
				Engineering Sciences}\ }\textbf {\bibinfo {volume} {370}},\ \bibinfo {pages}
		{2848} (\bibinfo {year} {2012})}\BibitemShut {NoStop}%
	\bibitem [{\citenamefont {Bentley}\ \emph {et~al.}(2016)\citenamefont
		{Bentley}, \citenamefont {Batista},\ and\ \citenamefont
		{Steinbock}}]{oliver03}%
	\BibitemOpen
	\bibfield  {author} {\bibinfo {author} {\bibfnamefont {M.~R.}\ \bibnamefont
			{Bentley}}, \bibinfo {author} {\bibfnamefont {B.~C.}\ \bibnamefont
			{Batista}}, \ and\ \bibinfo {author} {\bibfnamefont {O.}~\bibnamefont
			{Steinbock}},\ }\href {\doibase 10.1021/acs.jpca.6b03859} {\bibfield
		{journal} {\bibinfo  {journal} {The Journal of Physical Chemistry A}\
		}\textbf {\bibinfo {volume} {120}},\ \bibinfo {pages} {4294} (\bibinfo {year}
		{2016})},\ \bibinfo {note} {pMID: 27266993},\ \Eprint
	{http://arxiv.org/abs/http://dx.doi.org/10.1021/acs.jpca.6b03859}
	{http://dx.doi.org/10.1021/acs.jpca.6b03859} \BibitemShut {NoStop}%
	\bibitem [{\citenamefont {Rymer}\ and\ \citenamefont
		{Williams-Jones}(2000)}]{15}%
	\BibitemOpen
	\bibfield  {author} {\bibinfo {author} {\bibfnamefont {H.}~\bibnamefont
			{Rymer}}\ and\ \bibinfo {author} {\bibfnamefont {G.}~\bibnamefont
			{Williams-Jones}},\ }\href@noop {} {\bibfield  {journal} {\bibinfo  {journal}
			{Geophys. Res. Lett.}\ }\textbf {\bibinfo {volume} {27}},\ \bibinfo {pages}
		{2389} (\bibinfo {year} {2000})}\BibitemShut {NoStop}%
	\bibitem [{\citenamefont {Schneider}\ \emph {et~al.}(2012)\citenamefont
		{Schneider}, \citenamefont {Rasband},\ and\ \citenamefont {Eliceiri}}]{17}%
	\BibitemOpen
	\bibfield  {author} {\bibinfo {author} {\bibfnamefont {C.~A.}\ \bibnamefont
			{Schneider}}, \bibinfo {author} {\bibfnamefont {W.~S.}\ \bibnamefont
			{Rasband}}, \ and\ \bibinfo {author} {\bibfnamefont {K.~W.}\ \bibnamefont
			{Eliceiri}},\ }\href@noop {} {\bibfield  {journal} {\bibinfo  {journal}
			{Nature Methods}\ }\textbf {\bibinfo {volume} {9}},\ \bibinfo {pages} {671}
		(\bibinfo {year} {2012})}\BibitemShut {NoStop}%
	\bibitem [{\citenamefont {Karol}(2003)}]{Karol2003}%
	\BibitemOpen
	\bibfield  {author} {\bibinfo {author} {\bibfnamefont {R.~H.}\ \bibnamefont
			{Karol}},\ }\href@noop {} {\emph {\bibinfo {title} {{Chemical Grouting and
					Soil Stabilization}}}},\ \bibinfo {edition} {3rd}\ ed.\ (\bibinfo
	{publisher} {Marcel Dekker, Inc.},\ \bibinfo {year} {2003})\BibitemShut
	{NoStop}%
	\bibitem [{\citenamefont {Sumino}\ \emph {et~al.}(2007)\citenamefont {Sumino},
		\citenamefont {Kitahata}, \citenamefont {Seto},\ and\ \citenamefont
		{Yoshikawa}}]{sumino01}%
	\BibitemOpen
	\bibfield  {author} {\bibinfo {author} {\bibfnamefont {Y.}~\bibnamefont
			{Sumino}}, \bibinfo {author} {\bibfnamefont {H.}~\bibnamefont {Kitahata}},
		\bibinfo {author} {\bibfnamefont {H.}~\bibnamefont {Seto}}, \ and\ \bibinfo
		{author} {\bibfnamefont {K.}~\bibnamefont {Yoshikawa}},\ }\href {\doibase
		10.1103/PhysRevE.76.055202} {\bibfield  {journal} {\bibinfo  {journal}
			{Physical Review E: Statistical, Nonlinear, and Soft Matter Physics}\
		}\textbf {\bibinfo {volume} {76}},\ \bibinfo {pages} {055202} (\bibinfo
		{year} {2007})}\BibitemShut {NoStop}%
	\bibitem [{\citenamefont {Sumino}\ \emph {et~al.}(2016)\citenamefont {Sumino},
		\citenamefont {Yamada}, \citenamefont {Nagao}, \citenamefont {Honda},
		\citenamefont {Kitahata}, \citenamefont {Melnichenko},\ and\ \citenamefont
		{Seto}}]{sumino02}%
	\BibitemOpen
	\bibfield  {author} {\bibinfo {author} {\bibfnamefont {Y.}~\bibnamefont
			{Sumino}}, \bibinfo {author} {\bibfnamefont {N.~L.}\ \bibnamefont {Yamada}},
		\bibinfo {author} {\bibfnamefont {M.}~\bibnamefont {Nagao}}, \bibinfo
		{author} {\bibfnamefont {T.}~\bibnamefont {Honda}}, \bibinfo {author}
		{\bibfnamefont {H.}~\bibnamefont {Kitahata}}, \bibinfo {author}
		{\bibfnamefont {Y.~B.}\ \bibnamefont {Melnichenko}}, \ and\ \bibinfo {author}
		{\bibfnamefont {H.}~\bibnamefont {Seto}},\ }\href {\doibase
		10.1021/acs.langmuir.6b00107} {\bibfield  {journal} {\bibinfo  {journal}
			{Langmuir}\ }\textbf {\bibinfo {volume} {32}},\ \bibinfo {pages} {2891}
		(\bibinfo {year} {2016})},\ \bibinfo {note} {pMID: 26938640},\ \Eprint
	{http://arxiv.org/abs/http://dx.doi.org/10.1021/acs.langmuir.6b00107}
	{http://dx.doi.org/10.1021/acs.langmuir.6b00107} \BibitemShut {NoStop}%
	\bibitem [{\citenamefont {Haudin}\ \emph {et~al.}(2014)\citenamefont {Haudin},
		\citenamefont {Cartwright}, \citenamefont {Brau},\ and\ \citenamefont
		{De~Wit}}]{PNAS}%
	\BibitemOpen
	\bibfield  {author} {\bibinfo {author} {\bibfnamefont {F.}~\bibnamefont
			{Haudin}}, \bibinfo {author} {\bibfnamefont {J.~H.~E.}\ \bibnamefont
			{Cartwright}}, \bibinfo {author} {\bibfnamefont {F.}~\bibnamefont {Brau}}, \
		and\ \bibinfo {author} {\bibfnamefont {A.}~\bibnamefont {De~Wit}},\
	}\href@noop {} {\bibfield  {journal} {\bibinfo  {journal} {Proc. Natl. Acad.
			Sci. USA}\ }\textbf {\bibinfo {volume} {111}},\ \bibinfo {pages} {17363}
	(\bibinfo {year} {2014})}\BibitemShut {NoStop}%
\bibitem [{\citenamefont {Haudin}\ \emph {et~al.}(2015)\citenamefont {Haudin},
	\citenamefont {Cartwright},\ and\ \citenamefont {De~Wit}}]{JPCC}%
\BibitemOpen
\bibfield  {author} {\bibinfo {author} {\bibfnamefont {F.}~\bibnamefont
		{Haudin}}, \bibinfo {author} {\bibfnamefont {J.~H.~E.}\ \bibnamefont
		{Cartwright}}, \ and\ \bibinfo {author} {\bibfnamefont {A.}~\bibnamefont
		{De~Wit}},\ }\href@noop {} {\bibfield  {journal} {\bibinfo  {journal} {J.
			Phys. Chem. C}\ }\textbf {\bibinfo {volume} {119}},\ \bibinfo {pages} {15067}
	(\bibinfo {year} {2015})}\BibitemShut {NoStop}%
\bibitem [{\citenamefont {Haudin}\ and\ \citenamefont {De~Wit}(2015)}]{05}%
\BibitemOpen
\bibfield  {author} {\bibinfo {author} {\bibfnamefont {F.}~\bibnamefont
		{Haudin}}\ and\ \bibinfo {author} {\bibfnamefont {A.}~\bibnamefont
		{De~Wit}},\ }\href@noop {} {\bibfield  {journal} {\bibinfo  {journal} {Phys.
			of Fluids}\ }\textbf {\bibinfo {volume} {27}},\ \bibinfo {pages} {113101}
	(\bibinfo {year} {2015})}\BibitemShut {NoStop}%
\bibitem [{\citenamefont {Cartwright}\ \emph {et~al.}(2002)\citenamefont
	{Cartwright}, \citenamefont {Garcia-Ruiz}, \citenamefont {Novella},\ and\
	\citenamefont {Otalora}}]{08}%
\BibitemOpen
\bibfield  {author} {\bibinfo {author} {\bibfnamefont {J.~H.~E.}\
		\bibnamefont {Cartwright}}, \bibinfo {author} {\bibfnamefont {J.~M.}\
		\bibnamefont {Garcia-Ruiz}}, \bibinfo {author} {\bibfnamefont {M.~L.}\
		\bibnamefont {Novella}}, \ and\ \bibinfo {author} {\bibfnamefont
		{F.}~\bibnamefont {Otalora}},\ }\href@noop {} {\bibfield  {journal} {\bibinfo
		{journal} {J. Colloid Interface Sci.}\ }\textbf {\bibinfo {volume} {256}},\
	\bibinfo {pages} {351} (\bibinfo {year} {2002})}\BibitemShut {NoStop}%
\bibitem [{\citenamefont {Tabeling}\ \emph {et~al.}(1986)\citenamefont
	{Tabeling}, \citenamefont {Zocchi},\ and\ \citenamefont {Libchaber}}]{12}%
\BibitemOpen
\bibfield  {author} {\bibinfo {author} {\bibfnamefont {P.}~\bibnamefont
		{Tabeling}}, \bibinfo {author} {\bibfnamefont {G.}~\bibnamefont {Zocchi}}, \
	and\ \bibinfo {author} {\bibfnamefont {A.}~\bibnamefont {Libchaber}},\
}\href@noop {} {\bibfield  {journal} {\bibinfo  {journal} {J. Fluid Mech.}\
}\textbf {\bibinfo {volume} {177}},\ \bibinfo {pages} {67} (\bibinfo {year}
{1986})}\BibitemShut {NoStop}%
\bibitem [{\citenamefont {Park}\ and\ \citenamefont
	{Homsy}(1984)}]{ParkHomsy1984}%
\BibitemOpen
\bibfield  {author} {\bibinfo {author} {\bibfnamefont {C.-W.}\ \bibnamefont
		{Park}}\ and\ \bibinfo {author} {\bibfnamefont {G.~M.}\ \bibnamefont
		{Homsy}},\ }\href {\doibase 10.1017/S0022112084000367} {\bibfield  {journal}
	{\bibinfo  {journal} {Journal of Fluid Mechanics}\ }\textbf {\bibinfo
		{volume} {139}},\ \bibinfo {pages} {291} (\bibinfo {year}
	{1984})}\BibitemShut {NoStop}%
\bibitem [{\citenamefont {Cartwright}\ \emph {et~al.}(2011)\citenamefont
	{Cartwright}, \citenamefont {Escribano},\ and\ \citenamefont
	{Sainz-Diaz}}]{colour}%
\BibitemOpen
\bibfield  {author} {\bibinfo {author} {\bibfnamefont {J.~H.~E.}\
		\bibnamefont {Cartwright}}, \bibinfo {author} {\bibfnamefont
		{B.}~\bibnamefont {Escribano}}, \ and\ \bibinfo {author} {\bibfnamefont
		{C.~I.}\ \bibnamefont {Sainz-Diaz}},\ }\href {\doibase 10.1021/la104192y}
{\bibfield  {journal} {\bibinfo  {journal} {Langmuir}\ }\textbf {\bibinfo
		{volume} {27}},\ \bibinfo {pages} {3286} (\bibinfo {year} {2011})},\ \bibinfo
{note} {pMID: 21391635},\ \Eprint
{http://arxiv.org/abs/http://dx.doi.org/10.1021/la104192y}
{http://dx.doi.org/10.1021/la104192y} \BibitemShut {NoStop}%
\bibitem [{pre({\natexlab{a}})}]{pressure}%
\BibitemOpen
\href@noop {} {} \bibinfo {note} {The pressure drop from
	the pressure gauge to the injection point can be calculated based on the
	Hagen-Poiseuille equation as $\Delta p = 8 \eta L Q/(\pi r^4)$. In the case
	when $Q$ = 0.01, 3 and 100 mL/min., $\Delta p$ = 0.2, 60 and 2000 Pa,
	respectively.}\BibitemShut {Stop}%
\bibitem [{pre({\natexlab{b}})}]{pressure2}%
\BibitemOpen
\href@noop {} {} \bibinfo {note} {The pressure drop inside
	of the cell can be estimated by the fomular for Poiseulle flow where $-dp/dx
	= 12 \eta V$, if the pressure drop appears due to the radial flow of viscous
	liquid whose velocity is denoted by $V$. For radially symmetric outword flow
	at the distance $r$ from the center, $V=Q/2 \pi r h$. Therefore, $-dp/dx =
	6\eta Q/(\pi r h^3)$. Thus, the pressure drop within the Hele-Shaw cell is
	$\Delta p = -6 Q \eta \ln(R/r) / (\pi h^3)$, where $R$ and $r$ are the radius
	of the cell and the inlet, respectively. For $R$ = 50 mm, $r$ = 0.5 mm,
	$\ln(R/r) \sim$ 4.6. For $Q =$ 0.01, 3, and 10 mL/min., $\Delta p =$ 0.08, 3
	and 80 Pa, respectively.}\BibitemShut {Stop}%
\bibitem [{\citenamefont {Nakanishi}(2006)}]{02}%
\BibitemOpen
\bibfield  {author} {\bibinfo {author} {\bibfnamefont {H.}~\bibnamefont
		{Nakanishi}},\ }\href {\doibase 10.1103/PhysRevE.73.061603} {\bibfield
	{journal} {\bibinfo  {journal} {Phys. Rev. E}\ }\textbf {\bibinfo {volume}
		{73}},\ \bibinfo {pages} {061603} (\bibinfo {year} {2006})}\BibitemShut
{NoStop}%
\bibitem [{\citenamefont {Sumino}\ and\ \citenamefont {Wagatsuma}()}]{future1}%
\BibitemOpen
\bibfield  {author} {\bibinfo {author} {\bibfnamefont {Y.}~\bibnamefont
		{Sumino}}\ and\ \bibinfo {author} {\bibfnamefont {S.}~\bibnamefont
		{Wagatsuma}},\ }\href@noop {} {\ } \bibinfo {note} {in preparation} \BibitemShut {NoStop}%
\bibitem [{\citenamefont {Podgorski}\ \emph {et~al.}(2007)\citenamefont
	{Podgorski}, \citenamefont {Sostarecz}, \citenamefont {Zorman},\ and\
	\citenamefont {Belmonte}}]{Belmonte1}%
\BibitemOpen
\bibfield  {author} {\bibinfo {author} {\bibfnamefont {T.}~\bibnamefont
		{Podgorski}}, \bibinfo {author} {\bibfnamefont {M.~C.}\ \bibnamefont
		{Sostarecz}}, \bibinfo {author} {\bibfnamefont {S.}~\bibnamefont {Zorman}}, \
	and\ \bibinfo {author} {\bibfnamefont {A.}~\bibnamefont {Belmonte}},\
}\href@noop {} {\bibfield  {journal} {\bibinfo  {journal} {Phys. Rev. E}\
}\textbf {\bibinfo {volume} {76}},\ \bibinfo {pages} {016202} (\bibinfo
{year} {2007})}\BibitemShut {NoStop}%
\bibitem [{\citenamefont {Nagatsu}\ \emph {et~al.}(2008)\citenamefont
	{Nagatsu}, \citenamefont {Bae}, \citenamefont {Kato},\ and\ \citenamefont
	{Tada}}]{Nagatsu1}%
\BibitemOpen
\bibfield  {author} {\bibinfo {author} {\bibfnamefont {Y.}~\bibnamefont
		{Nagatsu}}, \bibinfo {author} {\bibfnamefont {S.-K.}\ \bibnamefont {Bae}},
	\bibinfo {author} {\bibfnamefont {Y.}~\bibnamefont {Kato}}, \ and\ \bibinfo
	{author} {\bibfnamefont {Y.}~\bibnamefont {Tada}},\ }\href {\doibase
	10.1103/PhysRevE.77.067302} {\bibfield  {journal} {\bibinfo  {journal} {Phys.
			Rev. E}\ }\textbf {\bibinfo {volume} {77}},\ \bibinfo {eid} {067302}
	(\bibinfo {year} {2008})}\BibitemShut {NoStop}%
\bibitem [{\citenamefont {Alberts}\ \emph {et~al.}(2008)\citenamefont
	{Alberts}, \citenamefont {Johnson}, \citenamefont {Lewis}, \citenamefont
	{Raff}, \citenamefont {Roberts},\ and\ \citenamefont {Walter}}]{cell}%
\BibitemOpen
\bibfield  {author} {\bibinfo {author} {\bibfnamefont {B.}~\bibnamefont
		{Alberts}}, \bibinfo {author} {\bibfnamefont {A.}~\bibnamefont {Johnson}},
	\bibinfo {author} {\bibfnamefont {J.}~\bibnamefont {Lewis}}, \bibinfo
	{author} {\bibfnamefont {M.}~\bibnamefont {Raff}}, \bibinfo {author}
	{\bibfnamefont {K.}~\bibnamefont {Roberts}}, \ and\ \bibinfo {author}
	{\bibfnamefont {P.}~\bibnamefont {Walter}},\ }\href
{http://www.worldcat.org/isbn/0815332181} {\emph {\bibinfo {title}
		{{Molecular Biology of the Cell, Fifth Edition}}}},\ \bibinfo {edition}
{5th}\ ed.\ (\bibinfo  {publisher} {Garland Science},\ \bibinfo {year}
{2008})\BibitemShut {NoStop}%
\bibitem [{\citenamefont {Rhee}(1977)}]{Rhee1977}%
\BibitemOpen
\bibfield  {author} {\bibinfo {author} {\bibfnamefont {S.~K.}\ \bibnamefont
		{Rhee}},\ }\href {\doibase 10.1007/BF00548176} {\bibfield  {journal}
	{\bibinfo  {journal} {Journal of Materials Science}\ }\textbf {\bibinfo
		{volume} {12}},\ \bibinfo {pages} {823} (\bibinfo {year} {1977})}\BibitemShut
{NoStop}%
\bibitem [{pre({\natexlab{c}})}]{suppl}%
\BibitemOpen
\href@noop {} {} \bibinfo {note} {See Supplemental Material at [URL will be inserted by publisher] for movies of patterns.}\BibitemShut {Stop}%

\end{thebibliography}
%

\end{document}